\documentclass[hyper]{JHEP3}

\input{epsf}
\usepackage{epsfig}
\usepackage{amssymb}
\usepackage{amsfonts}
\usepackage{amsbsy}
\usepackage{amsmath}

\usepackage{epstopdf}
\usepackage{graphics}      
\usepackage{graphicx}      
\numberwithin{equation}{section}

\newcommand{\met}{g_{\mu\nu}}

\title{Instability of Near-Extremal Black Holes in $\mathcal{N}=2$, $d=4$ Supergravity}

\author{Hyeyoun Chung\\
hyeyoun@physics.harvard.edu\\	
Jefferson Physical Laboratory, Harvard University,\\ 17 Oxford St., Cambridge, MA 02138, USA}\date{\small\today} 
\date{\small\today} 
\abstract{As a precursor to studying the bound states of multiple non-extremal black holes in $\mathcal{N}=2$, $d=4$ supergravity, we investigate the stability of a near-extremal D0-D4 black hole in the probe limit, when the parameters of the black hole solution lie within a certain regime. We determine whether it is possible to form bound states of this ``core'' non-extremal black hole with BPS probe particles, and whether it is possible for the ``core'' black hole to decay by the emission of such BPS probes either to a local minimum of the probe potential, or spatial infinity. We first carry out a qualitative analysis of the probe potential to determine when quantum tunneling of probes from the black hole is possible. We then find the wavefunction of the scattered probe by using the WKB approximation to solve the Dirac equation in the black hole background, and use this solution to compute the tunneling amplitude.} 
\begin{document}

\section{Introduction}

A class of black holes that has not been extensively studied so far consists of the non-extremal black hole solutions to $\mathcal{N}=2$, $d=4$ supergravity\cite{Gibbons, Ortin, HHGG}, that are a generalization of the well-known extremal (both BPS and non-BPS) ``attractor'' solutions\cite{Attractor}. These solutions consist of the background metric, together with a set of complex scalars and electromagnetic gauge fields, and can be completely characterized by the black hole's electric and magnetic charges, the parameter $c$ giving the deviation from extremality, and the value of the scalars at spatial infinity. In this work we study a particular class of these black hole solutions that have one electric charge $Q_0$, one magnetic charge $P_1$, and one complex scalar field. Thus these solutions can be specified by four independent quantities: $Q_0$, $P_1$, the extremality parameter $c$, and the value of the scalar field at spatial infinity. We package these quantities into two parameters, $\frac{1}{\tilde{c}_2}$ and $\frac{1}{\tilde{c}_4}$, that we use to classify the various regimes in which these solutions lie.

The BPS attractor solutions are known to exist in multi-centered configurations that are stable bound states\cite{DenefSUGRA, DenefBates}. These multicentered configurations should remain valid, stable solutions even when deformed away from extremality\cite{HHGG}, but fully backreacted, non-extremal, multicentered solutions have not yet been found. As the probe limit of BPS particles in a single-centered BPS background gives interesting insights into the fully backreacted supersymmetric multicentered solutions (such as the equilibrium distance between the centers)\cite{DenefSUGRA, DenefBates, VDThesis}, it is reasonable to assume that studying the behavior of BPS probes in the background of a single-centered non-extremal black hole should provide clues to the existence and stability of multi-centered non-extremal black hole solutions. 

We thus consider a central ``core'' non-extremal black hole, that can be surrounded by BPS probes (in a multi-centered configuration with a large black hole at the center and the remaining black holes being small enough relative to the ``core'' black hole that they can be treated as BPS probes), and study the static potential of the probes in the black hole background. We assume that the probes are small enough that mutual interactions between them can be ignored. If the probe potential has a local minimum whose free energy is lower than that of the free energy at the black hole horizon, then the probe can form a stable bound state with the background black hole\cite{HHGG, Chowd} (and when fully back-reacted, this could give a bound state of two black holes.) A single black hole whose charge is equal to the sum of the ``core'' charge and the probe charge is unstable to the emission of such a probe to form this bound state, as the probe can tunnel through the potential barrier to the local minimum of the potential. If the probe has a lower free energy at spatial infinity than it does at the black hole horizon, then the ``core'' black hole is unstable, as it can emit probes that can tunnel through the potential barrier to escape to infinity. We cannot predict the endpoint of this evolution, as at some point the probe approximation (where the interaction between the probes is negligible) will become inapplicable: however, it is possible that the ``core'' black hole will continue to emit probes to infinity until we are left with a hot dilute gas.

In this work we consider the emission of such charged probes from a certain class of non-extremal black holes in the near-extremal case, where the parameters describing the black hole solutions satisfy $c \lll \frac{1}{\tilde{c}_2}\sim \frac{1}{\tilde{c}_4}$. We would like to know if these black holes are unstable to the emission of charged BPS probes, and if so, to determine whether these probes can form bound states with the background black hole, and to compute the tunneling amplitudes for this emission process. We first study the qualitative features of the potential for a static charged probe in the black hole background, finding the maximum of the potential and the classical turning points. We also compute the rate for a charged particle to tunnel through the static potential between the classical turning points. This naive result for the tunneling rate through the potential barrier may also be applied to the case where the parameters of the black hole solution satisfy $c \lll \frac{1}{\tilde{c}_2}\lll \frac{1}{\tilde{c}_4}$.

We then find the wavefunction for a charged probe particle in the black hole background, by solving the curved space Dirac equation for this background. We use the WKB approximation to solve for the radial part of the wavefunction. We then compute the amplitude for a charged probe particle to be emitted from the black hole via quantum tunneling through the potential barrier, by calculating the ratio of the conserved current density at spatial infinity and at the black hole horizon.

This paper is organized as follows. In Section \ref{sec-Background} we introduce the non-extremal black hole solutions that we will be studying, together with the necessary formalism for describing these solutions. In Section \ref{sec-ProbeAction} we give the action for a charged probe particle in this black hole background, and carry out a qualitative analysis of the static probe potential. In Section \ref{sec-DiracEq} we set out the Dirac equation in the black hole background, and in Section \ref{sec-TunnAmp} we carry out the full tunneling analysis by solving the Dirac equation. We conclude in Section \ref{sec-Conclusion}.

\section{Non-extremal black holes in $\mathcal{N}=2$, $d=4$ supergravity}\label{sec-Background}

The action for the bosonic part of four-dimensional $\mathcal{N}=2$ supergravity coupled to massless vector multiplets takes the form:
\begin{align}
S_{4D} &= \frac{1}{16\pi}\int_{M_4} d^4 x \sqrt{-g}\left (R - 2G_{A\bar{B}}dz^A \wedge \star d\bar{z}^{\bar{B}} - F^I \wedge G_I\right),
\end{align}
where the $z^A$ $(A=1,\dots,n)$ are the vector multiplet scalars, the $F^I$ $(I=0,1,\dots,n)$ are the vector field strengths, the $G_I$ are the dual magnetic field strengths, and $G_{A\bar{B}} = \partial_A\partial_{\bar{B}}\mathcal{K}$ is derived from the Kahler potential
\begin{align}
\mathcal{K} = -\ln (i\int_X \Omega_0\wedge\bar{\Omega}_0)
\end{align}
where $\Omega_0$ is the holomophic 3-form on the Calabi-Yau manifold $X$. The normalized 3-form $\Omega = e^{\mathcal{K}/2}\Omega_0$.

The lattice of electric and magnetic charges $\Gamma$ is identified with $H^3(X,\mathbb{Z})$, the lattice of integral harmonic 3-forms on $X$. In the standard symplectic basis, a charge $\Gamma$ can be written as $\Gamma = (P^I, Q_I)$, with magnetic charges $P^I$ and electric charges $Q_I$. We can define a canonical, duality invariant, symplectic product $\langle, \rangle$ on the space of charges, which is given by:
\begin{align}\label{eq-IntProd}
\langle\Gamma, \tilde{\Gamma} \rangle = P^I\tilde{Q}_I - Q_I\tilde{P}^I
\end{align}
in the standard symplectic basis. The moduli-dependent central charge $Z(\Gamma,z)$ of $\Gamma$ is given by:
\begin{align}
Z(\Gamma,z) = -e^{\mathcal{K}/2}\langle \Gamma,\Omega\rangle
\end{align}
All the coefficients of the Lagrangian can be derived from a single prepotential $F(X)$, where the $X^A$ are projective coordinates such that $X^A = X^0z^A$ and $X^0$ is a gauge degree of freedom. In this paper we will consider prepotentials of the form:
\begin{align}\label{eq-PrePot}
F(X) = D\frac{(X^1)^3}{6X^0},
\end{align}
with $D=1$, so that we have one scalar field, $z$, and two electromagnetic vector potentials, $A^0$ and $A^1$ (together with their duals, $B^0$ and $B^1$.) In this case the moduli-dependent central charge for a charge $\gamma = (P^0, P^1, Q_1, Q_0)$ is given explicitly by:
\begin{align}
Z(\gamma,z) =\frac{\sqrt{3}}{2\sqrt{D(\mathrm{Im}z)^3}}\left (\frac{D}{6}P^0z^3 - \frac{D}{2}P^1z^2 + Q_1z + Q_0 \right )
\end{align}
This restriction still allows us to consider a large class of black hole solutions, as it has been shown that the general case with an arbitrary number of $n$ vector multiplets may be reduced to an effective theory with a single vector multiplet given by the prepotential (\ref{eq-PrePot}), by applying a suitable truncation\cite{DenefMoore}.

\subsection{The Black Hole Solution}

Non-extremal black hole solutions to this theory were first found in \cite{Gibbons} and \cite{Ortin}, and further studied in \cite{HHGG} and \cite{TunnelingString}. We consider the D0-D4 solutions of D0-charge $Q_0$ and D4-charge $P_1$, which may be described by a charge vector $\Gamma = (P^1, Q_0)$. In analogy with the well-known extremal black hole solutions\cite{Attractor}, a non-extremal solution can be given in terms of two functions $H_0, H_1$:
\begin{align}\label{eq-Harmonic}
H_0 \equiv \frac{|Q_0|}{c}\sinh(c\tau + c_2),\qquad H_1 \equiv \frac{|P_1|}{c}\sinh (c\tau + c_4)
\end{align}
where $c_2$ and $c_4$ are constants, $c$ denotes the deviation from extremality, and $\tau$ is an inverse radial coordinate such that $\tau\to\infty$ at the black hole horizon and $\tau\to 0$ at spatial infinity (in the extremal limit, the functions $H_0, H_1$ are harmonic.) If $Q_0$ and $P_1$ are both positive, then the solution is BPS in the extremal limit $c\to 0$. If $Q_0$ and $P_1$ have differing sign (where without loss of generality we can take $Q_0 < 0$), then the solution is non-BPS in the extremal limit. In this work we will restrict ourselves to the near-extremal regime where $c \lll \frac{1}{\tilde{c}_2}$ and $c \lll \frac{1}{\tilde{c}_4}$.

The black hole metric is given by\cite{Ortin, HHGG}:
\begin{align}
ds^2 = -e^{2U(\tau)}dt^2 + e^{-2U(\tau)}\left ( \frac{c^4}{\sinh^4 c\tau} d\tau^2 + \frac{c^2}{\sinh^2 c\tau}d\Omega_2^2\right )
\end{align}
where
\begin{align}
e^{-2U} &= \sqrt{\frac{2}{3}H_0H_1^3}.
\end{align}
The scalar field $z = iy$ is given by:
\begin{align}
y = \sqrt{\frac{6H_0}{H_1}},
\end{align}
and the electromagnetic vector potentials are given by:
\begin{align}
A^0 &= \frac{1}{2Q_0}\left ( \sqrt{c^2 + \frac{Q_0^2}{H_0^2}} - c\right ) dt,\qquad A^1 = P_1(1-\cos\theta)d\phi\nonumber\\
B_0 &= Q_0(1-\cos\theta)d\phi,\qquad B_1 = -\frac{3}{2P_1}\left ( \sqrt{c^2 + \frac{P_1^2}{H_1^2}} - c\right ) dt.
\end{align}
The integration constants $c_2$ and $c_4$ are given by:
\begin{align}
\sinh c_2 = \frac{cy_0^{3/2}}{2\sqrt{3}|Q_0|},\qquad \sinh c_4 = \frac{\sqrt{3}c}{|P_1|y_0^{1/2}},
\end{align}
where $y_0$ is the value of the scalar field at spatial infinity. Note that the solution is completely determined by the charges $(P^1, Q_0)$, the extremality parameter $c$, and the value $y_0$ of the scalar field at spatial infinity. Thus four independent parameters are needed to specify the solution. In the rest of this paper we will refer to the parameters $(P^1, Q_0)$, $c, c_2$, and $c_4$, but it should be kept in mind that one of these is redundant.

We will find it convenient to define the parameters:
\begin{align}
\tilde{c}_2 \equiv \frac{\sinh c_2}{c},\qquad \tilde{c}_4 \equiv \frac{\sinh c_4}{c}
\end{align}

The ADM mass $M$ of the black hole can be read off from the metric:
\begin{align}
M &= \frac{1}{4}\sqrt{c^2 + \frac{12Q_0^2}{y_0^3}} + \frac{3}{4}\sqrt{c^2 + \frac{P_1^2y_0}{3}}
\end{align}
where $y_0$ is the value of $y$ at spatial infinity. The entropy of the black hole is:
\begin{align}
S &= \pi \left (c + \sqrt{c^2 + \frac{12Q_0^2}{y_0^3}} \right )^{1/2} \left (c + \sqrt{c^2 + \frac{P_1^2y_0}{3}} \right )^{3/2}
\end{align}
and the Hawking temperature is:
\begin{align}\label{eq-THawking}
T_H = \frac{c}{2S}
\end{align}
In this work we will use the radial coordinate $r$ defined by:
\begin{align}
r \equiv \frac{c}{\sinh c\tau}
\end{align}
In these coordinates the metric becomes:
\begin{align}\label{eq-Metricr}
ds^2 = -e^{2U(r)}dt^2 + e^{-2U(r)}\left ( \frac{1}{\left ( 1+\frac{c^2}{r^2}\right)} dr^2 + r^2d\Omega_2^2\right )
\end{align}
And the functions determining the solutions are:
\begin{align}
H_0 &= |Q_0|\left (\frac{\cosh c_2}{r} + \tilde{c}_2 \sqrt{1+\frac{c^2}{r^2}} \right),\qquad\qquad H_1 = |P_1|\left (\frac{\cosh c_4}{r} + \tilde{c}_4 \sqrt{1+\frac{c^2}{r^2}} \right)
\end{align}

\section{The Probe Action}\label{sec-ProbeAction}

Before solving for the wavefunction of a BPS particle of charge $\gamma$ in the black hole background, it will be helpful to study the probe action for such a particle in this background, which is given by\cite{ProbeAction}:
\begin{align}\label{eq-ProbeAction}
S_{\gamma} = -\int \mu\,\mathrm{d}s - \frac{1}{2}\int \langle \gamma, \mathbb{A}_\mu \rangle \mathrm{d}x^\mu
\end{align}
where $\mu$ is the moduli-dependent mass of the particle, given by:
\begin{align}
\mu = |Z(\gamma,z)|
\end{align}
The \textit{static} probe action is:
\begin{align}
S_p &= -\int  \mu\,\sqrt{-g_{tt}}\,\mathrm{d}t - \frac{1}{2}\int \langle \gamma, \mathbb{A}_t \rangle \mathrm{d}t\\
&= \int V_p \,\mathrm{d}t\nonumber
\end{align}
where $V_p$ is the static probe potential. This can be written in the form $V_p = V_g + V_{\mathrm{em}}$, where $V_g = e^U|Z(\gamma,z)|$ is the mass term, given by\cite{HHGG}:
\begin{align}
V_g &= \frac{1}{4}\sqrt{\left (\frac{q_0}{H_0} + \frac{3p_1}{H_1} \right )^2 + \frac{6H_0}{H_1}\left ( \frac{q_1}{H_0} - \frac{p_0}{H_1}\right )^2},
\end{align}
and $V_{\mathrm{em}}$ is the electromagnetic coupling term, given by:
\begin{align}
V_{\mathrm{em}} &= -\frac{1}{4}\frac{q_0}{Q_0}\left ( \sqrt{c^2 + \frac{Q_0^2}{H_0^2}}-c\right ) - \frac{3}{4}\frac{p_1}{P_1}\left ( \sqrt{c^2 + \frac{P_1^2}{H_1^2}}-c\right ).
\end{align}
We can read off the electromagnetic vector potential for this configuration from the full (i.e. non-static) probe action (\ref{eq-ProbeAction}):
\begin{align}\label{eq-GaugePot}
A_t &= -\frac{1}{4}\frac{q_0}{Q_0}\left ( \sqrt{c^2 + \frac{Q_0^2}{H_0^2}} - c\right ) -\frac{3}{4}\frac{p_1}{P_1}\left ( \sqrt{c^2 + \frac{P_1^2}{H_1^2}} - c\right )\nonumber\\
A_\phi &= \frac{\langle\gamma,\Gamma\rangle}{2} (1-\cos\theta)
\end{align}

\subsection{Qualitative Analysis of the Probe Potential}\label{sec-ProbePot}

In this section we carry out a qualitative analysis of the static probe potential in the near-extremal regime $c \lll \frac{1}{\tilde{c}_2}, \frac{1}{\tilde{c}_4}$. We will find it useful to define the following quantities:
\begin{align}
A_1 &\equiv \frac{1}{4}\frac{q_0}{Q_0}+\frac{3}{4}\frac{p_1}{P_1}\label{eq-A1}\\
E_0 &\equiv \sqrt{\frac{2}{3}|Q_0||P_1|^3}\label{eq-E0}\\
\mu_0^2 &\equiv |Z(\gamma,z)|^2_{r \lll \frac{1}{\tilde{c}_2}} = \frac{E_0}{16}\left [\left (\frac{q_0}{|Q_0|} + \frac{3p_1}{|P_1|} \right )^2 + \frac{6|Q_0|}{|P_1|}\left (\frac{q_1}{|Q_0|} - \frac{p_0}{|P_1|} \right )^2 \right ]\label{eq-Mu0}\\
\mu_\infty^2 &\equiv |Z(\gamma,z)|^2_{r\to\infty} = \frac{1}{16}\left [\left (\frac{q_0}{\tilde{c}_2|Q_0|} + \frac{3p_1}{\tilde{c}_4|P_1|} \right )^2 + \frac{6|Q_0|\tilde{c}_2}{|P_1|\tilde{c}_4}\left (\frac{q_1}{\tilde{c}_2|Q_0|} - \frac{p_0}{\tilde{c}_4|P_1|} \right )^2 \right ]\label{eq-MuInf}
\end{align}
We will carry out a full scattering analysis only in the case where $\frac{1}{\tilde{c}_2} \sim \frac{1}{\tilde{c}_4}$, which allows the calculations to be simplified. Note that in this case we have $\mu_0^2 \sim \mu_\infty^2$, and $E_0 \sim \frac{1}{\tilde{c}_2^2}$. 

The probe has zero potential energy at the horizon $r=0$, while its potential energy at spatial infinity is
\begin{align}
V_p|_{r=\infty} &= \mu_\infty - \frac{1}{4}\frac{q_0}{Q_0\tilde{c}_2}-\frac{3}{4}\frac{p_1}{P_1\tilde{c}_4}+cA_1 \\
&= \mu_\infty - A_1(E_0^{\frac{1}{2}}-c)
\end{align}
Tunneling of a probe particle to spatial infinity is only possible if this potential energy is non-positive, so that $\mu_\infty \leq A_1(E_0^{\frac{1}{2}}-c)$ (note that $E_0^{\frac{1}{2}} \sim 1/\tilde{c}_2 >>> c$, so that in order for tunneling to spatial infinity to be allowed, we must have $A_1 > 0$.) We can see from (\ref{eq-GaugePot}) that the intersection product $\langle\gamma,\Gamma\rangle$ between the probe and the black hole determines the magnetic part of the vector potential, not the electrostatic part: so the sign of the intersection product is independent of the sign of the static potential at spatial infinity.

From (\ref{eq-A1})-(\ref{eq-E0}) and (\ref{eq-MuInf}) we see that if the black hole is BPS in the extremal limit, then $V_p|_{r=\infty} > 0$ unless the extremality parameter $c =0$ and the intersection product $\langle \gamma, \Gamma\rangle$ between the probe charge $\gamma$ and the black hole charge $\Gamma$ is zero, in which case $V_p|_{r=\infty} = 0$. In all other cases, the emission and absorption of probes is infinitely suppressed from black holes that are BPS in the extremal limit. Thus, apart from the special case of the emission of probe particles with $\langle \gamma, \Gamma\rangle = 0$ from BPS black holes, we are only considering the scattering of probes from black holes that are non-BPS in the extremal limit. This result is very natural, as the fully backreacted, two-centered BPS black hole solution has an angular momentum proportional to $\langle \Gamma_1, \Gamma_2\rangle$ where $\Gamma_1$ and $\Gamma_2$ are the charges of the two black holes, whereas a single-centered BPS black hole has zero angular momentum\cite{DenefSUGRA, DenefBates, VDThesis}. Thus, if $\langle \gamma, \Gamma\rangle \neq 0$, then emission of the probe $\gamma$ from the background black hole is forbidden by conservation of angular momentum in the extremal limit, when the background black hole is BPS. 

We can study the shape of $V_p$ in the two regions $r \lll \frac{1}{\tilde{c}_2}$ and $r \ggg c$. For $r \lll \frac{1}{\tilde{c}_2}$ we have:
\begin{align}\label{eq-Vp1}
V_p|_{r\lll \frac{1}{\tilde{c}_2}} = \frac{r}{E_0^{\frac{1}{2}}}\mu_0 - cA_1\left (\sqrt{1+\frac{r^2}{c^2}}-1 \right )
\end{align}
Solving $\frac{\partial V_p}{\partial r}|_{r=r_0} = 0$ gives:
\begin{align}
r_0 = c\sqrt{\frac{1}{\alpha_0^2}-1}
\end{align}
where
\begin{align}
\alpha_0^2 &\equiv 1 - \frac{\mu_0^2}{A_1^2E_0}.
\end{align}
Since $V''(r_0) < 0$, this is a maximum of the potential. And since $\mu_\infty \leq A_1(E_0^{\frac{1}{2}}-c)$ when emission to infinity is allowed, and $\mu_\infty^2 = \mu_0^2$, we have the following lower bound on $\alpha_0^2$ (remembering that we are in the near-extremal regime $c\tilde{c}_2 \lll 1$):
\begin{align}
\alpha_0^2 &\geq \frac{A_1^2E_0 - (A_1E_0^{\frac{1}{2}} - cA_1)^2}{A_1^2E_0}\\
&\sim c\tilde{c}_2
\end{align}
This also puts the following upper bound on $r_0$:
\begin{align}
r_0 \leq \frac{c^{\frac{1}{2}}}{\tilde{c}_2^{\frac{1}{2}}} \lll \frac{1}{\tilde{c}_2},
\end{align}
so we see that the maximum of the potential is indeed in the region $r \lll \frac{1}{\tilde{c}_2}$ in all the cases where emission to infinity is allowed.

In the region $r \ggg c$ we have:
\begin{align}
V_p|_{r\ggg c} = \frac{r(\tilde{c}_2\mu_\infty - A_1)}{1+r\tilde{c}_2} + cA_1
\end{align}
so that
\begin{align}
\frac{\partial V_p}{\partial r} = \frac{\mu_\infty E_0^{-\frac{1}{2}} - A_1}{(1+\tilde{c}_2r)^2},
\end{align}
which is $\leq 0$ in all the cases where emission to spatial infinity is allowed. Thus we see that with $\frac{1}{\tilde{c}_2} \sim \frac{1}{\tilde{c}_4}$, the potential has one maximum at $r=r_0$ in the region $r\sim c$, then decreases continuously, tending towards the constant value $V_p(\infty) = \mu_\infty - A_1(E_0^{\frac{1}{2}}-c)$ at spatial infinity. The potential does not have a local minimum, and thus the ``core'' black hole cannot form a bound state with a BPS probe. A probe that is emitted will tunnel through the potential barrier and escape to spatial infinity.

The maximum value of the potential at $r_0$ is given by
\begin{align}\label{eq-MaxE}
\epsilon_{\mathrm{max}} &\equiv V_p(r_0) = cA_1(1-\alpha_0)
\end{align}
Thus we can write the energy for a general scattered/emitted low energy particle as:
\begin{align}\label{eq-MaxEMod}
\epsilon = cA_1(1-\beta\alpha_0)
\end{align}
where $1 \leq \beta \leq \frac{1}{\alpha_0}$. The classical turning points for this particle are:
\begin{align}\label{eq-rTP}
r_{\pm} = \frac{c}{\alpha_0}\left [\beta\sqrt{1-\alpha_0^2} \pm \sqrt{\beta^2 - 1} \right ]
\end{align}
A low energy probe particle that has Poincare energy less than or equal to the maximum value $\epsilon_{\textrm{max}}$ of the potential can be emitted from the horizon, tunnel through the barrier between the classical turning points $r_-$ and $r_+$, and escape to spatial infinity. Alternatively, a particle may come in from infinity and scatter off the potential, either being reflected back from the barrier, or tunneling through from the turning point $r_+$ to $r_-$ before falling into the horizon. If the background black hole is BPS, then the static probe potential $V_p$ for a probe particle with $\langle \gamma, \Gamma\rangle = 0$ is actually flat for all $r$, so there is no barrier to emission or absorption.

Note that $V_p$ takes the same form as (\ref{eq-Vp1}) in the region $r \lll \frac{1}{\tilde{c}_2}$ when $\frac{1}{\tilde{c}_2} \lll \frac{1}{\tilde{c}_4}$. Thus, if $\alpha_0 \leq 1$, then the probe potential in this case also has a maximum at $r_0$ with the same value $\epsilon_{\mathrm{max}}$. Although a complete scattering analysis for regions $r \ggg c$ is more complicated in this case, the tunneling rate between the turning points $r_\pm$ of this potential is the same as in the case $\frac{1}{\tilde{c}_2} \sim \frac{1}{\tilde{c}_4}$ and may be calculated in the same manner, as we will see in Section \ref{sec-Gammat}.

\subsection{The Tunneling Rate Through The Probe Potential Barrier}\label{sec-Gammat}

Even without completing the full scattering calculation, we can obtain an estimate of the tunneling rate $\Gamma_t$ through the potential barrier, which is defined as:
\begin{align}
\Gamma_t \equiv e^{-\int_{r_-}^{r_+} |p_r'|dr'},
\end{align}
where $r_\pm$ are the classical turning points for the probe in the static potential, and $p_r$ is the radial canonical momentum, given by:
\begin{align}
p_r \equiv \frac{\partial\mathcal{L}}{\partial\dot{r}}.
\end{align}
The Lagrangian density $\mathcal{L}$ can be found from the probe action (\ref{eq-ProbeAction}):
\begin{align}
\mathcal{L} = -\mu\sqrt{e^{2U} - \left (\frac{e^{-2U}}{1+\frac{c^2}{r^2}}\right)\dot{r}^2} - A_t,
\end{align}
where $A_t$ is given by (\ref{eq-GaugePot}). This gives
\begin{align}
p_r = \frac{\mu e^{-2U}\dot{r}}{\sqrt{e^{2U}\left(1+\frac{c^2}{r^2}\right)^2 - e^{-2U}\left(1+\frac{c^2}{r^2}\right)\dot{r}^2}}
\end{align}
The conserved canonical energy $\epsilon$ satisfies:
\begin{align}
\epsilon &= \dot{r}\frac{\partial\mathcal{L}}{\partial\dot{r}} - \mathcal{L}\\
&= A_t + \sqrt{\left(1+\frac{c^2}{r^2}\right)e^{4U}p_r^2 + e^{2U}\mu^2}
\end{align}
Thus we find that
\begin{align}\label{eq-Pr}
p_r^2 &= \frac{e^{-4U}}{\left (1+\frac{c^2}{r^2}\right)}\left ((\epsilon - A_t)^2 - e^{2U}\mu^2 \right ).
\end{align}

For the range of parameters that we are considering for the background black hole, the classical turning points $r_\pm$ for the scattered probe particle always lie in the region $r \lll c$. Thus, defining the coordinate $z \equiv r^2/c^2$, the classical turning points lie in the region $z \lll 1$ and are given by (\ref{eq-rTP}):
\begin{align}
z_\pm &= \frac{\beta^2}{\alpha_0^2}\left (\pm 1+\sqrt{(1-\alpha_0^2)(1-\frac{1}{\beta^2})} \right )^2 - 1\\
z_+ - z_- &= \frac{4\beta}{\alpha_0^2}\sqrt{(1-\alpha_0^2)(\beta^2 - 1)}
\end{align}
We find that
\begin{align}
\lim_{\zeta\to 0} |\mathrm{Re}\left(i \left [p_r(z_+ - \zeta) - p_r(z_- - \zeta) \right ]\right)| = -a_1\alpha_0\pi(\beta-1),
\end{align}
and so
\begin{align}\label{eq-Gammat}
\Gamma_{\mathrm{t}} = e^{-a_1\alpha_0\pi(\beta-1)}
\end{align}
As mentioned at the end of Section \ref{sec-ProbePot}, $\Gamma_t$ also gives the tunneling rate through the potential barrier in the regime $\frac{1}{\tilde{c}_2}\lll \frac{1}{\tilde{c}_4}$, as the probe potential $V_p$ has the same form in the region $r \lll \frac{1}{\tilde{c}_2}$ as when $\frac{1}{\tilde{c}_2}\sim \frac{1}{\tilde{c}_4}$.

\section{The Dirac Equation}\label{sec-DiracEq}

In this section we carry out a full analysis of the scattering and emission of a charged probe of charge $\gamma$ from the black hole. We would like to solve for the wavefunction of a probe of Poincare energy $\epsilon$ scattering off the background black hole. In \cite{HallHalo}, it was found that the ground state wavefunction describing a light BPS probe particle of charge $\gamma$ in the background of another, heavy BPS particle of charge $\Gamma$, is given by a monopole spherical harmonic\cite{WuYang} corresponding to a configuration with total angular momentum $(\langle\gamma, \Gamma\rangle - 1)/2$. This can be thought of as the light BPS probe going into a spin-$1/2$ state aligned with the radial magnetic field of the background BPS particle, thus minimizing the energy of the configuration, and contributing one spin quantum opposite to the intrinsic field angular momentum, which is $\langle \gamma, \Gamma \rangle/2$. This is analogous to the problem of scattering a Dirac particle of charge $Ze$ in the background of a magnetic monopole of strength $g$\cite{WuYang, Yang2}, where the intersection product $\langle \gamma, \Gamma\rangle$ corresponds to the quantity $Zeg$. Since we are trying to solve the curved space version of this problem (though for higher energy states as well as the ground state), it is reasonable to assume that the wavefunction describing a probe particle emitted from the background black hole will obey the curved-space Dirac equation in the black hole background, and correspond to the probe going into a spin-$1/2$ state aligned with the radial magnetic field of the background black hole.

Using the vierbein formalism, the Dirac equation in curved space is:
\begin{align}\label{eq-DiracEq}
i\gamma^a V_a^\mu \partial_\mu\Psi + \frac{i}{2}\gamma^a V_a^\mu V_b^\nu V_{c\nu;\mu}\Sigma^{bc} \Psi - \gamma^aV_a^\mu A_\mu\Psi = \mu\Psi,
\end{align}
where $\mu$ is the mass of the particle given by $\mu = |Z(\gamma, z)|$, $A_\mu$ is the electromagnetic gauge potential, and $\Psi$ is a 4-component spinor. In our case $A_\mu$ is given by (\ref{eq-GaugePot}). Note that any choice of the vector potential $A_\phi$ around a magnetic monopole must have singularities. Thus, the specific form of $A_\phi$ given in (\ref{eq-GaugePot}) represents a gauge choice that is non-singular in some region $R_a$ (in this case, the region $r > 0$, $0 \leq \phi < 2\pi$, and $0 \leq \theta < \pi$.) In order to cover the entire space outside the magnetic monopole, we must divide the space into two regions and use a different gauge in each region. We can define the second region $R_b$ as $r > 0$, $0 \leq \phi < 2\pi$, and $0 < \theta \leq \pi$, and use the gauge choice
\begin{align}
A_\phi = -\frac{\langle\gamma,\Gamma\rangle}{2} (1+\cos\theta)
\end{align}
in this region.

Because of this fact, the components of the Dirac spinor
\begin{align}
\Psi &= \left ( \begin{array}{c}\psi_0\\ \psi_1\\ \psi_2\\ \psi_3\end{array}\right ),
\end{align}
in a magnetic monopole background are given by sections on a line bundle, not a function\cite{WuYang}. The angular part of the section can be expanded in monopole spherical harmonics $Y_{j,l,m}(\theta,\phi)$ characterized by the quantum numbers $(j,l,m)$, where
\begin{align}
j &= 0, \frac{1}{2}, 1,\dots\\
l &= |j|, |j|+1, |j|+2,\\
m &= -l, -l + 1, \dots, l,
\end{align}
and
\begin{align}\label{eq-MonHarm1}
Y_{j,l,m}(\theta,\phi) &= e^{i((m+j)\phi} \Theta_{j,l,m}(\theta)\quad\mbox{in region}\,\,R_a\\
Y_{j,l,m}(\theta,\phi) &= e^{i((m-j)\phi} \Theta_{j,l,m}(\theta)\quad\mbox{in region}\,\,R_b\label{eq-MonHarm2}
\end{align}
for the same function $\Theta_{j,l,m}(\theta)$. 

For a suitable choice of vierbeins (for details, see Appendix \ref{sec-App1}), we substitute the ansatz
\begin{align}\label{eq-Ansatz}
\psi_0 &=R_0(r)Y_{q+1/2,l,m}(\theta,\phi)e^{iq\phi}e^{-i\epsilon t}\\
\psi_1 &=R_1(r)Y_{q-1/2,l,m}(\theta,\phi)e^{iq\phi}e^{-i\epsilon t}\\
\psi_2 &=R_2(r)Y_{q+1/2,l,m}(\theta,\phi)e^{iq\phi}e^{-i\epsilon t}\\
\psi_3 &=R_3(r)Y_{q-1/2,l,m}(\theta,\phi)e^{iq\phi}e^{-i\epsilon t}
\end{align}
into (\ref{eq-DiracEq}), where we have defined
\begin{align}
q &\equiv \frac{\langle \gamma, \Gamma\rangle}{2}.
\end{align}
In the case when $q > 0$, there is a possible solution with $\psi_0 = \psi_2 =0$, and $l = q - \frac{1}{2}$. When $q < 0$, there is a possible solution with $\psi_1 = \psi_3 =0$, and $l = -q-\frac{1}{2}$. As expected, these solutions correspond to the probe particle being aligned with the radial magnetic field of the background black hole. The two cases are exactly analogous so from now on we will assume that $q < 0$. We obtain coupled radial equations for $R_0(r)$ and $R_2(r)$ of the form:
\begin{align}\label{eq-Radial}
&e^{-U}(\partial_t + iA_t+i\mu)R_0 + \frac{e^{\frac{3U}{2}}}{r}\partial_r(re^{-\frac{U}{2}}R_2) = 0\\
&e^{-U}(\partial_t + iA_t-i\mu)R_2 + \frac{e^{\frac{3U}{2}}}{r}\partial_r(re^{-\frac{U}{2}}R_0) = 0.
\end{align}

\subsection{The WKB Approximation}\label{sec-WKB}

We cannot solve the radial part of the Dirac equation exactly in this background. So in order to compute the tunneling amplitude, we will solve for the radial components of the wavefunction using the WKB approximation.

In order to apply the WKB approximation, we first substitute $q \to q/\hbar$ and $\epsilon \to \epsilon/\hbar$. Using the ansatz (\ref{eq-Ansatz}) and defining $T_{0,2}(r) \equiv re^{-U/2}R_{0,2}(r)$ gives the following equation for $T_0(r)$ (with an analogous equation for $T_2(r)$:
\begin{align}\label{eq-RadialT}
T_0'' &= -\frac{e^{-4U}}{\hbar^2\left (1+\frac{c^2}{r^2}\right)}\left ((\epsilon - A_t)^2 - e^{2U}\mu^2 \right )T_0\\
&\qquad\qquad + T_0'\frac{e^U\sqrt{1+c^2/r^2}}{(\epsilon e^{-U} - A_t e^{-U} + \mu)}\frac{d}{dr}\left [\frac{e^{-U}(\epsilon e^{-U} - A_t e^{-U} + \mu)}{\sqrt{1+c^2/r^2}} \right ]
\end{align}
where a prime denotes differentiation with respect to $r$. Substituting the ansatz $T_{0,2} = B_{0,2}e^{iS_{0,2}/\hbar}$ and taking terms to leading order in $1/\hbar$, we find:
\begin{align}\label{eq-S0}
S_{0,2}'^2 = \frac{e^{-4U}}{\left (1+\frac{c^2}{r^2}\right)}\left ((\epsilon - A_t)^2 - e^{2U}\mu^2 \right ).
\end{align}
Note that this is the same expression that gives the canonical momentum $p_r$ in (\ref{eq-Pr}). We need to go to the next order in $1/\hbar$ to find the equation for $B_0(r)$ (an exactly analogous equation gives $B_2(r)$):
\begin{align}\label{eq-B0}
B_0S_0'' + 2B_0'S_0' = \left [\frac{\frac{d}{dr}(\epsilon e^{-U} - A_t e^{-U} + \mu)}{(\epsilon e^{-U} - A_t e^{-U} + \mu)} + e^U\sqrt{1+c^2/r^2}\frac{d}{dr}\left (\frac{e^{-U}}{\sqrt{1+c^2/r^2}} \right) \right ]B_0S_0'
\end{align}
This next order in the WKB approximation is necessary in order to derive the connection formulae needed to extend the solution past the classical turning points, where the WKB approximation becomes invalid. The connection formulae are derived in Appendix \ref{sec-App2}. We can solve equations (\ref{eq-S0})-(\ref{eq-B0}) in the region $r \lll \frac{1}{\tilde{c}_2}$ (recall that both turning points lie in this region for the class of background black holes we are considering), and then patch this solution to the WKB solution in the region $c \lll r$. The details of the calculation are given in Appendix \ref{sec-App3}.

\subsection{The Tunneling Amplitude}\label{sec-TunnAmp}

In the regime $c \lll \frac{1}{\tilde{c}_2} \sim \frac{1}{\tilde{c}_4}$ we find that
\begin{align}
\cosh c_2,\cosh c_4 \approx 1\\
\sinh c_2 \approx c\tilde{c}_2 \lll 1
\end{align}
In order to calculate the tunneling amplitude for a probe particle, we want to find the equivalent of the probability density of the particle wavefunction at different values of $r$. In the case of a Dirac spinor, this is given by the time component of the conserved current density $J^\mu$, which is given by:
\begin{align}
J^\mu &= \bar{\Psi}\underline{\gamma}^\mu\Psi,\\
\nabla_\mu J^\mu &= 0,
\end{align}
where 
\begin{align}
\underline{\gamma}^\mu := V^\mu_a \gamma^a,\\
\bar{\Psi} := \Psi^\dagger \gamma^0,
\end{align}
and $\gamma^0$ indicates the flat-space gamma matrix.

For the background metric (\ref{eq-Metricr}), $J^0$ is given by:
\begin{align}\label{eq-FluxR}
J^0 &= e^{-U}(|R_0|^2 + |R_2|^2).
\end{align}
We now want to compute the ratio of charge densities at spatial infinity $r\to\infty$, and at the black hole horizon $r \to 0$. Given a conserved current $J^\mu$ satisfying $\nabla_\mu J^\mu = 0$, we can define the one-form $J_\mu = g_{\mu\nu} J^\nu$ and write the conservation condition as:
\begin{align}
\mathrm{d}(\star J) = 0
\end{align}
We then define the charge passing through a hypersurface $\mathcal{H}$ via:
\begin{align}
Q_{\mathcal{H}} = -\int_{\mathcal{H}} \star J.
\end{align}
We take $\mathcal{H}$ to be a hypersurface of constant time, $t$. We can then write:
\begin{align}
Q_{\mathcal{H}} = -\int_{\mathcal{H}} d^3 x \sqrt{|h|} n_\mu J^\mu
\end{align}
where $h_{ij}$ is the spatial metric and $n_\mu = \met n^\nu$ where $n^\nu$ is the normal vector to the hypersurface. We therefore have
\begin{align}
n^\nu &= (1,0,0,0)\\
n_\mu &= \left (-e^{2U},0,0,0 \right )\\
\sqrt{|h|} &=  \sqrt{\frac{e^{-6U}}{\left (1+c^2/r^2 \right )} r^4\sin^2\theta}.
\end{align}
We can then write:
\begin{align}
Q_{\mathcal{H}} &= \int_{\mathcal{H}} d^3 x \sqrt{|h|}\, e^{2U}\, J^0\\
&= \int_{\mathcal{H}} dr\, d\theta\, d\phi\, e^{U}\sqrt{\frac{e^{-6U}}{\left (1+c^2/r^2 \right )} r^4\sin^2\theta}\,\,(|R_0|^2 + |R_2|^2)\\
&= \int_{\mathcal{H}} dr\, d\theta\, d\phi\, \frac{e^{-2U} r^2\sin\theta}{\sqrt{1+c^2/r^2}} \,\,(|R_0|^2 + |R_2|^2).
\end{align}
It follows that to get the charge density $q(r)$ at $r$, we should integrate just over $\theta,\phi$:
\begin{align}
q(r) &= \int_{\mathcal{H}} d\theta\, d\phi\, \frac{e^{-2U} r^2\sin\theta}{\sqrt{1+c^2/r^2}} \,\,(|R_0|^2 + |R_2|^2)\\
&\sim \frac{e^{-2U} r^2}{\sqrt{1+c^2/r^2}}(|R_0|^2 + |R_2|^2)
\end{align}

We can now calculate the charge densities at the black hole horizon, and at spatial infinity, using the form of the Dirac spinor found in Appendix \ref{sec-App3}. We impose boundary conditions so that the wavefunction is entirely outgoing at $r\to\infty$. In order to obtain the emission amplitude, we should compute the charge density at $r\to 0$ using only the outgoing component at the black hole horizon. We find that:
\begin{align}
q(r)_{r\to\infty} &= r^2(|R_0|^2 + |R_2|^2) = |\tilde{B}_0|^2\\ 
q(r)_{r\to 0} &= \frac{E_0r}{c}(|R_0|^2 + |R_2|^2) = \frac{\alpha_0}{c^2(1-\beta\alpha_0)}|\tilde{B}_0|^2\left (\Gamma_t-\frac{1}{2\Gamma_t} \right )^2
\end{align}
for some constant $\tilde{B}_0$, where the tunneling rate $\Gamma_t$ is given by (\ref{eq-Gammat}). We can obtain the tunneling amplitude by taking the ratio of the charge densities:
\begin{align}
\frac{q(r)_{r\to\infty}}{q(r)_{r\to 0}} &= \frac{c^2(1-\beta\alpha_0)}{\alpha_0}\frac{1}{\left (\Gamma_t-\frac{1}{2\Gamma_t} \right )^2}\\
&= \frac{c^2(1-\beta\alpha_0)}{\alpha_0}\frac{1}{\left (e^{-a_1\alpha_0\pi(\beta-1)} - \frac{1}{2}e^{a_1\alpha_0\pi(\beta-1)} \right )^2}
\end{align}
We can see that the tunneling amplitude depends on the parameter $\alpha_0$, which is related to the difference between the gravitational and electrostatic parts of the static probe potential, as well as the energy of the probe (which is related to $(1-\beta\alpha_0)$). It also appears that the tunneling amplitude decreases to zero as the background black hole becomes extremal.

\section{Conclusion}\label{sec-Conclusion}

In this paper we have initiated a study of the stability of a class of non-extremal black holes in $\mathcal{N}=2$, $d=4$ supergravity. Our results provide some clues as to how such a black hole perturbed away from extremality might evolve over time. If the black hole is BPS in the extremal limit, then for the range of parameters we have considered, then it can only emit probe particles $\gamma$ such that the intersection product $\langle \gamma,\Gamma\rangle$ between the probe charge and the black hole charge is zero. The static potential felt by such a probe is flat, so that there is no barrier to emission or absorption. Thus, we would expect that such a black hole would decay by the emission of these probes to spatial infinity until it became extremal.

If the black hole is non-BPS in the extremal limit, then it can emit probes of different charges. We found that the decay rate increases as the parameter $\alpha_0^2$ decreases, where $\alpha_0^2$ is related to the difference between the gravitational and electrostatic parts of the static probe potential. Thus, we expect probes with small $\alpha_0^2$ to be emitted preferentially by the black hole. For the class of black holes we studied, we found that the probe potential has a single maximum at $r=r_0$, and decreases steadily as $r \to \infty$. Thus, as there is no local minimum in the probe potential, we expect that any particles emitted from the black hole will simply be ejected out to spatial infinity.

There are many possible routes for further investigation: it would be useful to extend these results to arbitrary values of the parameters $c, \frac{1}{\tilde{c}_2}$, and $\frac{1}{\tilde{c}_4}$, so that we are no longer restricted to the near-extremal limit. Another natural extension of our results would be to study the scattering of fermionic particles in the black hole background, and to compare the relative emission rates of different types of particles from the black hole.

\section{Acknowledgements}

I would like to thank Jacob Barandes, Frederik Denef, and Hajar Ebrahim for encouraging me that this idea was worth pursuing. I would also like to thank Clay Cordova, David Simmons-Duffin, and Xi Yin for helpful discussions. This work was funded in part by a Research Assistantship from Harvard's Center for the Fundamental Laws of Nature.

\begin{appendix}

\section{The Dirac Equation in the Vierbein Formalism}\label{sec-App1}

Here we give the explicit form of the Dirac equation for a particular choice of vierbein. We use the following representation for the $\gamma$-matrices in the Dirac equation (\ref{eq-DiracEq}):
\begin{align}
\gamma^0 &= \left ( \begin{array}{cc} 1 & 0\\ 0 & -1\end{array}\right )\hspace{2cm} \gamma^i = \left ( \begin{array}{cc} 0 & -\sigma^i\\ \sigma^i & 0\end{array}\right )
\end{align}
and the following representation for $\Sigma^{ab}$:
\begin{align}
\Sigma^{ab} &= -\frac{1}{4}[\gamma^a, \gamma^b]
\end{align}
We use the following choice of vierbein:
\begin{align}
V^0 &= e^{U}\,dt\\
V^1 &= e^{-U}r \, d\theta\\
V^2 &= e^{-U}r \sin\theta \, d\phi\\
V^3 &= e^{-U}\sqrt{1+\frac{c^2}{r^2}}\,dr
\end{align}
The Dirac equation for a spinor
\begin{align}
\Psi &= \left ( \begin{array}{c}\psi_0\\ \psi_1\\ \psi_2\\ \psi_3\end{array}\right )
\end{align}
then becomes:
\begin{align}
&e^{-U}(\partial_t + iA_t+i\mu)\psi_0 + \frac{e^{\frac{3U}{2}}}{r}\partial_r(re^{-\frac{U}{2}}\psi_2)\sqrt{1+\frac{c^2}{r^2}}\\
&\hspace{1cm}+ \frac{e^U}{r}\left (\partial_\theta - \frac{i}{\sin\theta}\partial_\phi - \left (q-\frac{1}{2} \right )\cot\theta + \frac{q}{\sin\theta} \right)\psi_3 = 0\\
&e^{-U}(\partial_t + iA_t+i\mu)\psi_1 - \frac{e^{\frac{3U}{2}}}{r}\partial_r(re^{-\frac{U}{2}}\psi_3)\sqrt{1+\frac{c^2}{r^2}}\\
&\hspace{1cm}+ \frac{e^U}{r}\left (\partial_\theta + \frac{i}{\sin\theta}\partial_\phi + \left (q+\frac{1}{2} \right )\cot\theta - \frac{q}{\sin\theta} \right)\psi_2 = 0\\
&e^{-U}(\partial_t + iA_t-i\mu)\psi_2 + \frac{e^{\frac{3U}{2}}}{r}\partial_r(re^{-\frac{U}{2}}\psi_0)\sqrt{1+\frac{c^2}{r^2}}\\
&\hspace{1cm}+ \frac{e^U}{r}\left (\partial_\theta - \frac{i}{\sin\theta}\partial_\phi - \left (q-\frac{1}{2} \right )\cot\theta + \frac{q}{\sin\theta} \right)\psi_1 = 0\\
&e^{-U}(\partial_t + iA_t-i\mu)\psi_3 - \frac{e^{\frac{3U}{2}}}{r}\partial_r(re^{-\frac{U}{2}}\psi_1)\sqrt{1+\frac{c^2}{r^2}}\\
&\hspace{1cm}+ \frac{e^U}{r}\left (\partial_\theta - \frac{i}{\sin\theta}\partial_\phi + \left (q+\frac{1}{2} \right )\cot\theta - \frac{q}{\sin\theta} \right)\psi_0 = 0
\end{align}
where we have defined
\begin{align}
q &\equiv \frac{\langle \gamma, \Gamma\rangle}{2}.
\end{align}
We now use the fact that
\begin{align}
\mathcal{D} &\equiv -\partial_\theta - \frac{i}{\sin\theta}\partial_\phi + j\cot\theta\\
\bar{\mathcal{D}} &\equiv -\partial_\theta + \frac{i}{\sin\theta}\partial_\phi - j\cot\theta
\end{align}
are raising and lowering operators for the monopole spherical harmonics that satisfy:
\begin{align}
\mathcal{D}Y_{j,l,m} &= \sqrt{(l-j)(l+j+1)}Y_{j+1,l,m}\\
\bar{\mathcal{D}}Y_{j,l,m} &= -\sqrt{(l+j)(l-j+1)}Y_{j-1,l,m}
\end{align}
We substitute the ansatz:
\begin{align}
\psi_0 &=R_0(r)Y_{q+1/2,l,m}(\theta,\phi)e^{iq\phi}e^{-i\epsilon t}\\
\psi_1 &=R_1(r)Y_{q-1/2,l,m}(\theta,\phi)e^{iq\phi}e^{-i\epsilon t}\\
\psi_2 &=R_2(r)Y_{q+1/2,l,m}(\theta,\phi)e^{iq\phi}e^{-i\epsilon t}\\
\psi_3 &=R_3(r)Y_{q-1/2,l,m}(\theta,\phi)e^{iq\phi}e^{-i\epsilon t},
\end{align}
and see that if $q > 0$, then there is a possible set of solutions to the Dirac equation with $\psi_0 = \psi_2 = 0$, and $l = q - \frac{1}{2}$. Similarly, if $q < 0$, then there is a possible set of solutions with $\psi_1 = \psi_3 = 0$, and $l = -q-\frac{1}{2}$.

\section{Solving the Dirac Equation Using the WKB Approximation}\label{sec-App3}

Here we give the details of the calculations for solving the Dirac Equation using the WKB approximation. As shown in Section \ref{sec-WKB}, this comes down to solving equations (\ref{eq-S0}) and (\ref{eq-B0}) for the radial parts of the Dirac spinor components.

In the region $r \ggg c$, Eq. (\ref{eq-S0}) becomes:
\begin{align}
S_0'^2 = E_0A_1^2\alpha_0^2\left (1 + \frac{E_0^{\frac{1}{2}}}{r} \right )^2
\end{align}
and Eq. (\ref{eq-B0}) becomes:
\begin{align}
r\left (1+\frac{r}{E_0^{1/2}} \right )\left (B_0S_0'' + 2B_0'S_0'\right) + B_0S_0' = 0
\end{align}
These equations can be solved exactly to give:
\begin{align}
B_0(r) &= \tilde{B}_0\\
S_0(r) &= \pm \frac{a_1\alpha_0}{E_0^{\frac{1}{2}}}(r+E_0^{\frac{1}{2}}\log r)\label{eq-WKBRFar2App}
\end{align}
As $r \to \infty$, the solutions become:
\begin{align}
\frac{e^{U/2}}{r}B_0(r) &= \frac{\tilde{B}_0}{r}\\
S_\pm(r) &= \pm \frac{a_1\alpha_0}{E_0^{\frac{1}{2}}}r
\end{align}
Imposing outgoing boundary conditions at infinity (so we only keep $S_+$), the far region solution is:
\begin{align}
\psi &\sim \frac{\tilde{B}_0}{r}e^{ia_1\alpha_0E_0^{-\frac{1}{2}}r}
\end{align}
In order to patch the solutions in the regions $r \ggg c$ and $r \lll \frac{1}{\tilde{c}_2}$, we evaluate $B_0(r)$ and $S_+(r)$ in the limit $r\lll \frac{1}{\tilde{c}_2}$, which gives:
\begin{align}
\frac{e^{U/2}}{r}B_0(r) &= \frac{\tilde{B}_0}{r^{\frac{1}{2}}E_0^{\frac{1}{4}}}\\
S_+(r) &= a_1\alpha_0\log r
\end{align}
Thus the solution in the region $r \ggg c$ is:
\begin{align}\label{eq-OutgoingSol}
\psi &\sim \frac{\tilde{B}_0}{r^{\frac{1}{2}}E_0^{\frac{1}{4}}}r^{ia_1\alpha_0}
\end{align}

Defining $z \equiv \frac{r^2}{c^2}$, in the region $r \lll \frac{1}{\tilde{c}_2}$, Eq. (\ref{eq-S0}) becomes:
\begin{align}\label{eq-S0z}
S_0'^2 = \frac{a_1^2}{4z^2(1+z)}\left ((\beta\alpha_0 - \sqrt{1+z})^2 +(\alpha_0^2-1)z \right ),
\end{align}
where a prime denotes differentiation with respect to $z$. And Eq. (\ref{eq-B0}) becomes:
\begin{align}
&2(1+z)B_0(S_0' + 2zS_0'') + 2z\left [1 - 2(1+z)\left (\frac{z^{1/2}\frac{d}{dz}(z^{-1/2}(\beta\alpha_0 - \sqrt{1+z}))}{\beta\alpha_0 - \sqrt{1+z} - z^{1/2}\sqrt{1-\alpha_0^2}}\right )\right]B_0S_0'\\
&\qquad+ 8z(1+z)B_0'S_0' = 0
\end{align}
Equation (\ref{eq-S0z}) can be solved exactly to give:
\begin{align}\label{eq-WKBSolRApp}
S_0(z) &= S_\pm(z)\\
&= \pm\frac{a_1}{2}\Biggl [-2\beta\alpha_0\mathrm{Arctanh}(\sqrt{1+z}) - \log(-z) +2\alpha_0\log\left[-\beta+\alpha_0\sqrt{1+z} + f(z)\right]\nonumber\\
&\qquad\qquad + (1-\beta\alpha_0)\log\left [ 1-f(z)+\alpha_0 \left ( \alpha_0(-1+\sqrt{1+z}) + \beta^2\alpha_0 -\beta(1+\sqrt{1+z} - f(z))\right )\right ]\nonumber\\
&\qquad\qquad + (1+\beta\alpha_0)\log\left [ 1+f(z)+\alpha_0 \left ( -\alpha_0(1+\sqrt{1+z}) + \beta^2\alpha_0 +\beta(1-\sqrt{1+z} + f(z))\right ) \right ] \Biggr]
\end{align}
where
\begin{align}
f(z) \equiv \sqrt{1-2\beta\alpha_0\sqrt{1+z} + (z+\beta^2)\alpha_0^2},
\end{align}
The form of $B_0(z)$ may also be determined exactly, but we will not give it here as it is extremely complicated and not particularly illuminating.

In order to patch the solution to the region $r \ggg c$, we solve Eq. (\ref{eq-S0})-(\ref{eq-B0}) for $z \ggg 1$:
\begin{align}
&S_0'^2 = \frac{a_1^2\alpha_0^2}{4z^2}\\
&4zB_0(S_0' + zS_0'') + 8z^2B_0'S_0' = 0
\end{align}
to give:
\begin{align}
\frac{e^{U/2}}{r}B_0(z) &\sim \frac{1}{z^{\frac{1}{4}}}\\
S_{\pm}(z) &= \pm\frac{a_1\alpha_0}{2}\log z
\end{align}
Matching to (\ref{eq-OutgoingSol}), we find that the solution in the region $z \ggg 1$ is:
\begin{align}\label{eq-WKBSolZggg1App}
\psi &\sim \frac{\tilde{B}_0c^{ia_1\alpha_0}}{E_0^{\frac{1}{4}}c^{\frac{1}{2}}z^{\frac{1}{4}}} z^{i\frac{a_1\alpha_0}{2}}
\end{align}
We can extend this solution up to the outer turning point at $z=z_+$, by taking $S_+(z)$ to select the wave that gives this form of $\psi$ for $z \ggg 1$. At the turning point, the WKB approximation becomes invalid, as $B_0(z)$ diverges. Thus, in order to continue the WKB solution past $z_+$, we must derive connection formulas that allow us to patch together the solutions in the regions $z < z_-$ and $z > z_-$. We then repeat the procedure at the other turning point $z_-$. This can be done using Airy functions: detailed calculations are given in Appendix \ref{sec-App2}.

Using the connection formulas, we find that the solution in the region $z < z_-$ is:
\begin{align}\label{eq-WKBSmallApp}
\psi &\sim \frac{c^{ia_1\alpha_0}}{E_0^{\frac{1}{4}}c^{\frac{1}{2}}} B_0(z) \left ( \left (\Gamma_t - \frac{1}{2\Gamma_t} \right )e^{iS_+(z)} - i\left ( \Gamma_t + \frac{1}{2\Gamma_t}\right )e^{iS_-(z)}\right )
\end{align}
where the tunneling amplitude $\Gamma_t$ is given by (\ref{eq-Gammat}). The functions $B_0(z)$ and $S_\pm(z)$ in the region $z \lll 1$ are determined by the equations:
\begin{align}
&S_0'^2 = \frac{a_1^2}{4z^2}(1-\beta\alpha_0)^2\\
&4B_0(S_0' + zS_0'') + 8zB_0'S_0' = 0,
\end{align}
which give:
\begin{align}
B_0(z) &=\frac{\alpha_0^{\frac{1}{2}}}{(1-\beta\alpha_0)^{\frac{1}{2}}}\frac{c^{ia_1\alpha_0}}{E_0^{\frac{1}{4}}c^{\frac{1}{2}}}\tilde{B}_0\\
S_0(z)_\pm &= \pm\frac{a_1}{2}(1-\beta\alpha_0)\log z
\end{align}
where $\tilde{B}_0$ is a constant. The wavefunction in the region $z \lll 1$ is thus:
\begin{align}
\psi &\sim \frac{e^{U/2}}{r}B_0e^{iS_0/\hbar}\\
&= \frac{\alpha_0^{\frac{1}{2}}}{(1-\beta\alpha_0)^{\frac{1}{2}}}\frac{c^{ia_1\alpha_0}}{E_0^{\frac{1}{2}}c}\tilde{B}_0 z^{-1/4}\left ( \left (\Gamma_t - \frac{1}{2\Gamma_t} \right )z^{+\frac{ia_1}{2}(1-\beta\alpha_0)} -i \left (\Gamma_t + \frac{1}{2\Gamma_t} \right )z^{-\frac{ia_1}{2}(1-\beta\alpha_0)}\right )
\end{align}
It may appear that the solution blows up and becomes unphysical as $z \to 0$, but as we will see in Section (\ref{sec-TunnAmp}), this is not a problem as the charge density always remains finite. This solution is the limit of the general solution for $z \lll 1$.

\section{Deriving the connection formulas for the WKB approximation using Airy functions}\label{sec-App2}

Here we outline the calculation for deriving the connection formulas that allow the WKB solutions to be extended across the turning points at $z_+$ and $z_-$. The general procedure is to solve radial equation using a linear approximation to the potential at the turning point, and then match this solution to the WKB approximation to the left and right of the turning point.

If we define the potential
\begin{align}
V(z) &\equiv \frac{a_1^2}{4z^2(1+z)}\left[ \left(\beta\alpha_0 - \sqrt{1+z} \right)^2 + (\alpha_0^2-1)z\right],
\end{align}
then the radial equations for the WKB ansatz $\psi = B_r(r)e^{\frac{i}{\hbar}(S(r) + \epsilon t)}Y_{q,l,m}(\theta,\phi)$ are:
\begin{align}
&S'^2 = V(z)\\
&2(1+z)B_r(S' + 2zS'') + 2z\left [1 - 2(1+z)\left (\frac{z^{1/2}\frac{d}{dz}(z^{-1/2}(\beta\alpha_0 - \sqrt{1+z}))}{\beta\alpha_0 - \sqrt{1+z} - z^{1/2}\sqrt{1-\alpha_0^2}}\right )\right]B_rS'\\
&\qquad+ 8z(1+z)B_r'S' = 0
\end{align}
We can rewrite the second equation as:
\begin{align}
2(1+z)B_r(S' + 2zS'') + 8z(1+z)B_r'S' + 2z\left [1 - 2(1+z)M_1(z,\alpha_0,\beta)\right]B_rS' = 0
\end{align}
for some function $M_1(z,\alpha_0,\beta)$.

The exact radial equation, given by (\ref{eq-RadialT}), is:
\begin{align}
T'' + M(z,\alpha_0,\beta)T' + V(z) T = 0
\end{align}
for some function $M(z,\alpha_0,\beta)$, where $T(r) \equiv re^{-U/2}R(r)$ and $R(r)$ is the radial part of the Dirac spinor components.

\subsection{The Linear Approximation}

Consider a turning point at $z=z_0$. Defining the coordinate $w \equiv z - z_0$, we can take the linear approximation to the potential around $z_0$:
\begin{align}
V(z) &\approx V(z_0) + V'(z_0)w\\
&= \rho w
\end{align}
where $\rho \equiv V'(z_0)$. Note that we have $\rho < 0$ for $z_0 = z_-$ and $\rho > 0$ for $z_0 = z_+$. The radial equation can then be approximated as:
\begin{align}
T''(w) + M(z_0,\alpha_0,\beta)T'(w) + \rho w T(w) = 0
\end{align}
This equation has the general solution:
\begin{align}
T(w) = e^{\frac{-wM(z_0,\alpha_0,\beta)}{2}}\left [C_1\mathrm{Ai}(w') + C_2\mathrm{Bi}(w') \right ]
\end{align}
where $\mathrm{Ai}(w')$ and $\mathrm{Bi}(w')$ are Airy functions, and we have defined
\begin{align}
w' \equiv (-\rho)^{\frac{1}{3}}w + \frac{M(z_0,\alpha_0,\beta)^2}{4(-\rho)^{\frac{2}{3}}}
\end{align}
At each turning point we assume that there is a region in which the linear approximation is valid, and that the second term in the expression for $w'$ is negligible compared to the first, so that we can take:
\begin{align}\label{eq-WPrime}
w' \approx (-\rho)^\frac{1}{3}w.
\end{align}

\subsection{The turning point at $z_+$}

First consider the turning point at $z=z_+$. The general solution in the region around $z_+$ where the linear approximation is valid, is:
\begin{align}\label{eq-LinearSol2}
e^{\frac{-wM(z_+,\alpha_0,\beta)}{2}}\left [\tilde{C}_1\mathrm{Ai}(w') + \tilde{C}_2\mathrm{Bi}(w') \right ]
\end{align}
where $w'$ is defined as in (\ref{eq-WPrime}). Since $\rho$ is positive, we find that
\begin{align}
w' \sim (-1)^\frac{1}{3}\rho^{\frac{1}{3}}w,
\end{align}
and thus in order to match the above solutions to the WKB approximation to the left and right of $z_+$, we need the asymptotic behavior of the Airy functions for $w' \to (-1)^\frac{1}{3} \times -\infty$:
\begin{align}
\mathrm{Ai}(w') &\sim \frac{1}{2\sqrt{\pi}(-w')^{\frac{1}{4}}}\left [-i e^{\frac{2i}{3}(-w')^{3/2}} + e^{\frac{-2i}{3}(-w')^{3/2}}\right]\\
\mathrm{Bi}(w') &\sim \frac{1}{2\sqrt{\pi}(-w')^{\frac{1}{4}}}\left [-i e^{\frac{-2i}{3}(-w')^{3/2}}+ e^{\frac{2i}{3}(-w')^{3/2}}\right]
\end{align}
and $w' \to (-1)^\frac{1}{3} \times \infty$:
\begin{align}
\mathrm{Ai}(w') &\sim \frac{1}{2\sqrt{\pi}w'^{\frac{1}{4}}}e^{\frac{2}{3}w'^{3/2}}\\
\mathrm{Bi}(w') &\sim \frac{1}{\sqrt{\pi}w'^{\frac{1}{4}}}\left [\frac{i}{2} e^{\frac{2}{3}w'^{3/2}} + e^{-\frac{2}{3}w'^{3/2}}\right]
\end{align}
We now need to solve the WKB equations to the left and right of the turning point. The radial equations for the WKB ansatz in the region where the linear approximation is valid are:
\begin{align}
&S'(w)^2 = \rho w\\
&2(1+z_+)B_r(S' + 2z_+S'') + 8z_+(1+z_+)B_r'S' + 2z_+\left [1 - 2(1+z_+)M_1(z_+,\alpha_0,\beta)\right]B_rS' = 0
\end{align}
These equations have the solutions:
\begin{align}
S(w) = \pm\frac{2i}{3}\sqrt{\rho w^3}
\end{align}
and
\begin{align}
B_r = \frac{B_0 e^{-\frac{wM(z_+,\alpha,\beta)}{2}}}{w^{\frac{1}{4}}},
\end{align}
where $B_0$ is a constant and $M(z_+,\alpha,\beta)$ is the same function as in (\ref{eq-LinearSol}). Note that $\rho w^3$ is negative on the left of the turning point, and positive on the right.

Since we are imposing the boundary condition that only the outgoing wave is present at spatial infinity, the WKB solution to the right of the turning point is:
\begin{align}
\frac{B_0 e^{-\frac{wM(z_+,\alpha,\beta)}{2}}}{w^{\frac{1}{4}}}e^{\frac{2i}{3}\sqrt{\rho w^3}}
\end{align}
In order to match the linear solution in (\ref{eq-LinearSol2}) to the WKB solution, note that:
\begin{align}
\frac{2}{3}(w')^{\frac{3}{2}} = \frac{2i}{3}(\rho w^3)^{\frac{1}{2}}
\end{align}
and use the asymptotic form of the Airy functions as $w'\to (-1)^{\frac{1}{3}}\times\infty$. We assume that there is a region in which the linear approximation is valid, and $|w'|$ is large enough for the linear solution to be well approximated by this asymptotic form. We find that:
\begin{align}
\tilde{C}_1 &= 2\sqrt{\pi}(-1)^\frac{1}{12}\rho^\frac{1}{12}B_0\\
\tilde{C}_2 &= 0.
\end{align}
We now want to match the linear solution to the WKB solution on the left of the turning point. In this region, the general WKB solution is:
\begin{align}
\frac{e^{-\frac{wM(z_+,\alpha,\beta)}{2}}}{w^{\frac{1}{4}}}\left [\tilde{D}_1e^{-\frac{2\sqrt{\rho}}{3}w^{3/2}}+\tilde{D}_2e^{\frac{2\sqrt{\rho}}{3}w^{3/2}} \right ]
\end{align}
The linear solution takes the following asymptotic form as $w' \to (-1)^\frac{1}{3} \times -\infty$:
\begin{align}
\frac{\tilde{C}_1 e^{-\frac{wM(z_0,\alpha,\beta)}{2}}}{2\sqrt{\pi}(-w')^{\frac{1}{4}}}\left [-i e^{\frac{2i}{3}(-w')^{3/2}} + e^{\frac{-2i}{3}(-w')^{3/2}}\right]
\end{align}
where
\begin{align}
\frac{2}{3}(-w')^{\frac{3}{2}} = \frac{2i}{3}(\rho)^{\frac{1}{2}}(-w)^{\frac{3}{2}}
\end{align}
Matching the linear solution to the WKB solution to the left of the turning point gives:
\begin{align}\label{eq-DCoeff}
\tilde{D}_1 &= -i\tilde{D}_2\\
&= -ie^{-i\pi/4}B_0.
\end{align}

\subsection{The turning point at $z_-$}

We now repeat the derivation of the connection formulas at $z= z_-$. The general solution in the region around $z_-$ where the linear approximation is valid is:
\begin{align}\label{eq-LinearSol}
e^{\frac{-wM(z_-,\alpha_0,\beta)}{2}}\left [C_1\mathrm{Ai}(w') + C_2\mathrm{Bi}(w') \right ],
\end{align}
As $\rho$ is now negative, and $w' \sim (-\rho)^\frac{1}{3}w$, we need the asymptotic behavior of the Airy functions for $w' \to -\infty$:
\begin{align}
\mathrm{Ai}(w') &\sim \frac{1}{2i\sqrt{\pi}(-w')^{\frac{1}{4}}}\left [e^{\frac{i\pi}{4}}e^{\frac{2i}{3}(-w')^{3/2}}- e^{\frac{-i\pi}{4}}e^{\frac{-2i}{3}(-w')^{3/2}}\right]\\
\mathrm{Bi}(w') &\sim \frac{1}{2\sqrt{\pi}(-w')^{\frac{1}{4}}}\left [e^{\frac{i\pi}{4}}e^{\frac{2i}{3}(-w')^{3/2}}+ e^{\frac{-i\pi}{4}}e^{\frac{-2i}{3}(-w')^{3/2}}\right]
\end{align}
and for $w' \to \infty$:
\begin{align}
\mathrm{Ai}(w') &\sim \frac{1}{2\sqrt{\pi}w'^{\frac{1}{4}}}e^{-\frac{2}{3}w'^{3/2}}\\
\mathrm{Bi}(w') &\sim \frac{1}{\sqrt{\pi}w'^{\frac{1}{4}}}e^{\frac{2}{3}w'^{3/2}}
\end{align}
The WKB solution to the right of the turning point is:
\begin{align}\label{eq-WKBD}
\frac{e^{-\frac{wM(z_-,\alpha,\beta)}{2}}}{w^{\frac{1}{4}}}\left [D_1e^{\frac{2\sqrt{|\rho|}}{3}w^{3/2}} + D_2e^{-\frac{2\sqrt{|\rho|}}{3}w^{3/2}} \right ]
\end{align}
where $D_1$ and $D_2$ are related to $\tilde{D}_1$ and $\tilde{D}_2$ in (\ref{eq-DCoeff}) by:
\begin{align}
\tilde{D}_1 &= \Gamma_{\mathrm{t}}D_1\\
\tilde{D}_2 &= \frac{1}{\Gamma_{\mathrm{t}}}D_2
\end{align}
\begin{align}
\Gamma_{\mathrm{t}} \equiv e^{-\int_{z_-}^{z_+} |R'(z')|dz'}
\end{align}
On the left hand side of the turning point, the general WKB solution is:
\begin{align}
\frac{e^{-\frac{wM(z_0,\alpha,\beta)}{2}}}{w^{\frac{1}{4}}}\left [G_1 e^{-\frac{2i}{3}\sqrt{\rho w^3}} + G_2 e^{\frac{2i}{3}\sqrt{\rho w^3}}\right]
\end{align}
Matching the linear and WKB solutions to the left and right of the turning point as before, gives:
\begin{align}
G_1 &= -iB_0\left(\Gamma_t + \frac{1}{2\Gamma_t}\right)\\
G_2 &= B_0\left(\Gamma_t - \frac{1}{2\Gamma_t}\right).
\end{align}
Thus, using these formulas we can extend the WKB solution across the turning points, from the region $z \ggg 1$ to the region $z \lll 1$.

\end{appendix}

\end{document}